# Kiladze Caldera: A Possible Cryovolcano on Pluto

A. Emran[1], D. P. Cruikshank[2], C. J. Ahrens[3], J. M. Moore[4], and O. L. White[5,4]
[1] NASA Jet Propulsion Laboratory, California Institute of Technology, Pasadena, CA 91109, USA; al.emran@jpl.nasa.gov
[2] Department of Physics, University of Central Florida, Orlando, FL 32816, USA
[3] NASA Goddard Space Flight Center, Greenbelt, MD 20707, USA
[4] NASA Ames Research Center, Moffett Field, CA 94035, USA
[5] SETI Institute, Mountain View, CA 94043, USA

## Abstract

In contrast with regional primarily methane composition, Kiladze and its surroundings exhibit a water-ice spectral signature that carries an ammoniated compound, similar to two other cryovolcanic sites on Pluto. The faulted structure of Kiladze, including shaping by numerous collapse pits and the distortion of the shape of the depression, are compatible with the surroundings in Hayabusa Terra, east of Sputnik Planitia. They are further compatible with an interpretation as a caldera formed during an era of an active cryovolcanic period that appears to be significantly more recent than the overall age of the planet's surface, possibly in the last several million years. In view of the size of the caldera and the large scale of the surrounding distribution of water ice, we suggest that Kiladze may have been a cryovolcano, in which one or more explosive events may have erupted ~1000 km$^3$ of icy cryomagma onto the surface.

## 1. Introduction

Images and spectroscopy of Pluto obtained by the New Horizons spacecraft that flew by the planet in 2015 July revealed several surface geologic structures and compositional units indicative of a variety of volcanic episodes (e.g., J. C. Cook et al. 2016; P. M. Schenk et al. 2018; O. L. White et al. 2021). These have been interpreted as manifestations of cryovolcanism, consistent with the volatile-rich character of Pluto as an outer solar system body and the planet's measured surface temperature (~40 K). Two structures are found in Pluto's heavily fractured Viking Terra, northwest of Sputnik Planitia (Figure 1), a basin containing a large expanse of convecting frozen nitrogen ($N_2$). Both exhibit graben, in which the defining fractures appear to have served as conduits along which a water-rich cryolava has debouched, flooded the graben floor and adjacent landscape structures, and then frozen. These cryolava-covered surfaces exhibit, in addition to the spectral signature of water, a weak spectral signature of ammonia, presumably in an ammoniated salt or some other ammonia compound resistant to Pluto's environment. Pure ammonia is readily destroyed by solar UV and solar wind particles, as well as by Galactic cosmic rays (D. P. Cruikshank et al. 2019b). The presence of an ammoniated compound is of interest in considerations of the origin and the conveyance of a water-rich cryomagma to Pluto's surface because of the role of ammonia in depressing the freezing point of water. Depending on the concentration of ammonia, this depression can amount to several tens of degrees, and up to 100° at the eutectic. Ammonia could thus be critical in maintaining a layer or pockets of fluid in Pluto's interior beyond the normal timescale of the cooling of the once molten core and mantle (F. Nimmo & W. B. McKinnon 2021). Hence, its discovery at sites of putative cryovolcanism is highly relevant. We note below that the NH*[6] signature is also found in Kiladze[7] and its surroundings from LEISA spectral data (A. Emran et al. 2023a).

The first exposure of cryolava identified is at the graben complex called Virgil Fossae (Figure 1(a)), where the cryolava is not only found in a winding channel on the floor of the main graben, but is also dispersed toward the south in a pattern identified as cryoclastic in nature (D. P. Cruikshank et al. 2019a; C. M. Dalle Ore et al. 2019). In addition, the cryolava in this region bears a distinct red-orange color. D. P. Cruikshank et al. (2019a) proposed that the red-orange pigment is organic in nature, and results from chemical reactions within the cryomagma in the interior, where warm water interacts with minerals and organic materials in Pluto's rocky interior. Laboratory experiments relevant to organic chemical reactions describe the formation of amino acids and other compounds, enabled and accelerated by the presence of ammonia in the fluid (e.g., D. P. Cruikshank et al. 2019a, 2020b). Indeed, pressurization of the fluid by ammonia in the source reservoir beneath the surface may have helped open a conduit to the surface (C. R. Martin & R. P. Binzel 2021). The second cryovolcanic feature in Viking Terra (Figure 1(a)) is a flooded crater and graben called Hardie and Uncama Fossa (D. P. Cruikshank et al. 2021). Here, the cryolava carries the $NH_3^*$ spectral signature, and also has a red-orange tint, but is not as intensely colored as that in Virgil Fossae.

K. N. Singer et al. (2022) describe a region of hummocky terrain southwest of Sputnik Planitia consisting of a mix of large mounds separated by large depressions. Two conspicuous and very broad features roughly circular in outline are Wright Mons and Piccard Mons, which are ~4–5 km and ~7 km in height and exhibit central pits. K. N. Singer et al. (2022) propose that these are cryovolcanic constructs built by eruptive episodes, although they do not have the spectral signature of the water ice that is characteristic of the cryovolcanic features at Virgil Fossae and Viking Terra. The apparent absence of these components may be a consequence of the unreliability of spectral data so near the terminator. Still, the steep topography of the mounds suggests that water ice must be present in some degree to support the kilometer-scale relief. Conversely, A. D. Howard et al. (2023) propose a scenario for the formation of Wright Mons and other nearby structures by the combination of the condensation of methane-rich and other endogenic gases from the subsurface through faults and their precipitation onto the landscape. In this view, the resulting structures are subsequently sculpted by solar radiation into the ridges characteristic of this region. In sum, these diverse structures, and possibly others so far unidentified, point to a picture of Pluto's interior and surface evolution producing a variety of surface expressions from previous epochs in the planet's history, leaving many unexplained clues.

[6] NH* denotes the presence of the ammonia molecule in some unknown form, probably a salt or a hydrate.
[7] Previously known as Pulfrich crater.

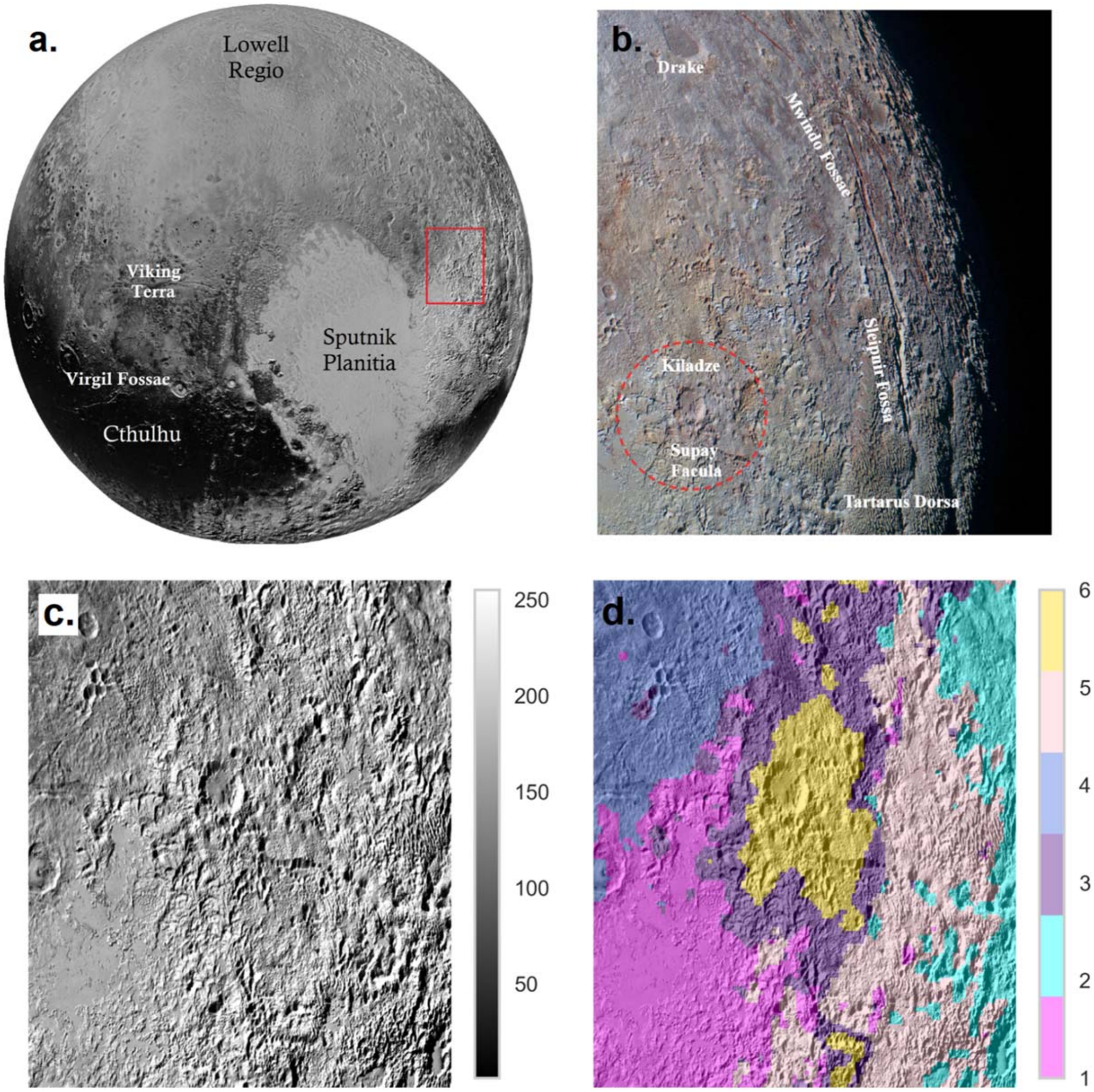

**Figure 1.** (a), (b) The global disk of Pluto showing major regions and the location of Kiladze (red outline). (c) The highest-resolution image of the region around Kiladze crater. The color-coded map (d) is derived from the clustering analysis by A. Emran et al. (2023a) with permission. Different colors denote regions with common spectral signatures; the yellow unit (#6) is dominated by water-ice spectral bands. The spectral properties of the remaining surface units (1–5) are found in A. Emran et al. (2023a). The lower panel's two figures (c) and (d) are recreated from A. Emran et al. (2023a). For details of the clustering technique, see the Appendix.

## 2. Geological Setting and Structure of Kiladze

In this paper, we investigate a major structure in Hayabusa Terra east of Sputnik Planitia that displays a degraded, rim-enclosed structure some 44 by 36 km in size, sitting in heavily pitted terrain. Kiladze (latitude $28\overset{\circ}{.}4$N, longitude $212\overset{\circ}{.}9$E) and its nearby surroundings first drew attention in global spectral maps (e.g., J. C. Cook et al. 2016; W. M. Grundy et al. 2016; S. Protopapa et al. 2017; B. Schmitt et al. 2017; A. Emran et al. 2023b), showing that its surface is dominated by water ice (Figure 1). This is in contrast to the regional surface composition, which is dominated spectroscopically by methane ice. Figure 1 shows the locational context of the area (a) and an image of Kiladze and its surroundings (b and c) for comparison with the same region color coded for the water ice (in yellow) in the corresponding image obtained with the LEISA spectrometer on New Horizons (d). For details about the way the ammoniated water ice was distinguished from other spectral features (largely methane) over the region, see A. Emran et al. (2023a) and the Appendix. We note that, while the yellow-coded region in Figure 1(d) delineates the contiguous area around the crater where ammoniated water ice dominates the spectrum (Figure A1), it is likely that exposures of water ice below the spatial resolution limit of LEISA occur well beyond its margins. One or more cataclysmic explosive eruptions that resulted in the excavation and collapse of what is seen as the Kiladze caldera would be expected to scatter the water-ice cryomagma widely for a thousand or more kilometers, leaving exposures too small to be seen in the data at hand. Such a wide dispersal of the ice could account for a substantial volume of the cryolava having erupted in one or more events. Figure 1(b) shows the distribution of a reddish color in Kiladze itself and some other structures, notably faults and pits chains. The base map image (Figure 1(c)) also shows a smooth texture covering most of the floor on Kiladze and a few additional irregular patches of smooth terrain about 100 km east (to the right) of the crater. There is also a conical mound of smooth terrain near the center of Kiladze, which in the transects (described below) is seen to stand several hundred meters above the mean level of the smooth terrain (see Section 4).

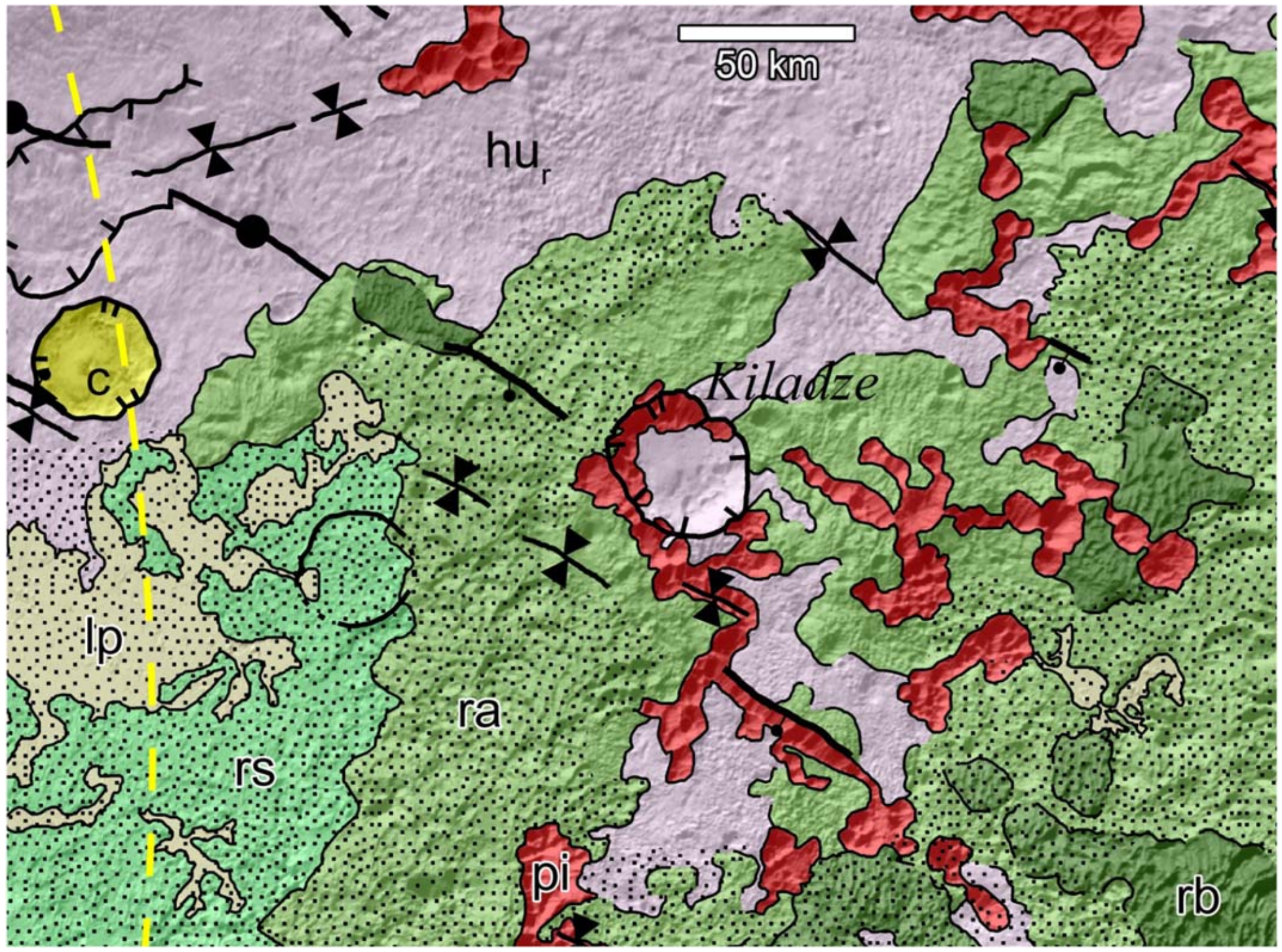

**Figure 2.** A portion of the 1:7,000,000 scale geological map of Pluto (O. L. White et al. 2022), centered on Kiladze at 212.°9E, 28.°4N. North is up. The rim of the crater is mapped with a depressed margin. Black lineations indicate fault lines. The dotted area indicates high-albedo terrain of East Tombaugh Regio. The yellow dashed line indicates the approximate crest of the broad annular ridge that contains the Sputnik impact basin. Unit labels are as follows: $hu_r$ = rough, undulating highland material; ra = ridged arcuate material; rb = bladed arcuate material; rs = subdued ridged material; lp = lowland plains material; c = crater material; pi = pit material.

This region is heavily fractured with normal faults, troughs, and graben that commonly exhibit a northwest–southeast orientation that is shared by most faults in this region of Pluto east of Sputnik Planitia (Figure 2; O. L. White et al. 2022). The rim of Kiladze is mapped with floor material consisting of rough, undulating highland material that is mapped across most of Hayabusa Terra to the north, with some collapse having taken place in the northern and western parts of the depression to form pit material. Chains of collapse pits east and south of the depression are aligned parallel to the orientation of nearby faults, suggesting that tectonism in this area is frequently associated with surface collapse (A. D. Howard et al. 2017). Kiladze itself forms an elliptical depression that is 46 km long by 36 km wide (giving an aspect ratio of 1.28), with the long axis aligned parallel to the northwest–southeast-oriented structural trends in the region. Surrounding Kiladze is the arcuate ridged material that, along with the bladed ridged material seen to the southeast that overlies it, is interpreted to be an accumulation of methane ice deposits, reaching 5 km thick, which were emplaced within the last few billion years of Pluto's history (J. M. Moore et al. 2016, 2021). Kiladze appears to crosscut the arcuate ridged material, implying that its formation postdates the emplacement of the methane deposits. However, because methane deposition dominates at higher, colder elevations on Pluto, there is the possibility that the arcuate ridged material may have been deposited after Kiladze formed, but accumulating only up to the rim of the depression and not on its steep walls or floor, which are typically depressed by 2–3 km below the surroundings. Models of seasonal ice (e.g., $N_2$, $CH_4$, and CO) sublimation and migration of the resulting gas indicate that much of this region is experiencing ongoing deposition of bright volatile ices that delineate East Tombaugh Regio, with deposited nitrogen ice ponding in depressions to form expanses of lowland plains such as those seen to the west (e.g., L. A. Young et al. 2021). Erosion of the ridged materials by flow of the nitrogen ice has modified them to form the subdued ridged material, which shows lower topographic relief than the arcuate or bladed ridged materials (e.g., L. A. Young et al. 2021).

A. Emran et al. (2023a) have compared the Kiladze complex to two features west of Sputnik Planitia where evidence of cryovolcanism has been revealed in the form of water ice that erupted onto the surface along fractures in Virgil Fossae and Uncama Fossa. The spectral signature of an unidentified ammoniated compound, similar to that in the other two sites, has also been found in the water ice of Kiladze (A. Emran et al. 2023a), but is not seen elsewhere on Pluto. Spectral modeling has also revealed a red-colored pigment in the Kiladze area, as reported by A. Emran et al. (2023a), a finding consistent with a visual map from MVIC color mosaic data (Figure 1(b)). However, the study by A. Emran et al. (2023a) leaves as an open question the source of the ammoniated water signature at Kiladze. As noted, the morphology of Kiladze is different from the geological structures (graben) observed in the other two areas (Virgil Fossae and Viking Terra) where the ammoniated water component was previously identified.

## 3. Age and Evolution of the Surfaces near Kiladze

The reflectance spectrum of the Kiladze area obtained by New Horizons is dominated by strong absorption bands of water and weak methane bands in the 1–2.5 $\mu$m region (A. Emran et al. 2023a). While other ices (e.g., $N_2$ and CO) occur elsewhere on Pluto, they are not clearly seen in the Kiladze area. An additional surface component, presumed to be distributed all over Pluto, is a layer of haze particles, formed photochemically in the atmosphere and precipitated across the surface over time. The formation of haze particles in Pluto's atmosphere is driven by complex molecules originating from a combination of gases and energy sources (e.g., K. E. Mandt et al. 2021). W. Grundy et al. (2018) reviewed issues of the timescale of formation of haze particles and settling to the surface, noting that there are uncertainties in both. In spite of these uncertainties, in view of the inventory of observed and calculated atmospheric photochemical products, averaged over Pluto's seasons, an estimated $350\,g\,cm^{-2}\,Gy^{-1}$ should condense. Once the individual particles accrete on one another to form larger particles of 1 $\mu$m or greater in size, they should all fall to the surface. If particle porosity is neglected and the average particle density is $1\,g\,cm^{-3}$, the surface would be coated to a thickness of 3.5 m in 1 Gy, or $\sim$14 m over the age of the solar system. Scaling this rate, a 10 $\mu$m thick layer of these particles, presumed to be optically thick, would accumulate in roughly 10 Pluto years of 248 Earth years each, or $\sim$2500 yr total.

Not all regions of the surface are of equal age, so the net accumulation is less where geological processes have occurred since the time of planet formation. W. Grundy et al. (2018) considered three geologically distinct regions on the hemisphere of Pluto observed with New Horizons, noting the differences in colors, compositions, and presumed amounts of aerosol particle accumulation. In order to explain the great diversity of albedo and color across Pluto's surface, some kind of surface chemical reactions, or an implausible spatial nonuniformity of formation and deposition (or removal, e.g., by varying degrees of scouring winds) must be invoked. Volatile surface ices sublimate and reprecipitate on various timescales and in different latitude zones (T. Bertrand et al. 2020), causing a redistribution of all the volatile ices, while water ice is not volatile at Pluto's temperature and is presumed to be a rigid bedrock over most or all of the planet. At the sites of cryovolcanism at Virgil Fossae and Viking Terra (D. P. Cruikshank et al. 2020a, 2019b), the water-rich cryolava is understood to have emerged from reservoirs in the subsurface.

The water-ice-rich surface at Kiladze raises the question of how it can remain exposed if it is continuously accumulating precipitating haze particles. While the Kiladze spectrum shows methane ice in addition to water ice, the methane can be explained as a seasonal condensation that comes and goes on some timescale associated with Pluto's axial inclination and orbital motion. In addition, there is a weak red coloration in the ices in and around the Kiladze structure, although it is much less strong than the reddish color in the Virgil Fossae and the Viking Terra exposures of cryovolcanic activity. This weaker and more widespread tint may arise from the fallout of atmospheric haze particles. In consideration of the discussion above, a logical conclusion is that the water ice at Kiladze is geologically young. If a surface accumulation of haze particles amounting to a layer of 10 mm is sufficient to mask the principal water-ice absorption bands, such a layer will form in $\sim$3 million years. If a thicker layer of aerosol particles would be needed to mask the ice spectral signature of the surface, the timescale for accumulation would be correspondingly longer, but surely much less than 1 Gy.

## 4. Kiladze Morphology

Kiladze is a heavily degraded rim-enclosed depression with multiple substructures in the rim, with pits of similar depth, all giving the impression of collapses, in the surroundings. The digital elevation model (DEM; P. M. Schenk et al. 2018) and a triangulated irregular network (TIN; W. R. Franklin 1973) product are shown in Figure 3, together with a perspective visualization of both models. The TIN product was generated from the DEM (P. M. Schenk et al. 2018) and was used to visually highlight the bottom of the depression. Most of the depression floor is smooth, and it has a rise near the center (Figures 3 and 4(a)). A smaller depression interrupts the southwest rim. Kiladze does not resemble other craters west of Sputnik Planitia, interpreted to be complex impact craters (see the discussion below).

In many respects, Kiladze resembles large calderas found on Earth and Mars. Terrestrial examples are the Yellowstone caldera, Long Valley Caldera, and Valles Caldera, all of which are considered resurgent calderas shaped by several major episodes of violent eruptive activity. A Martian example may be the collapsed pit craters at Noctis Labyrinthus (e.g., C. L. Kling et al. 2021). Similar structurally collapsed features have been reported within the Arabia Terra region on Mars, where several irregularly shaped craters resemble supervolcanic caldera constructs. These features are termed "plains-style caldera complexes" (J. R. Michalski and J. E. Bleacher 2013). These collapsed structures on Mars are linked to volcanic activity, more specifically supervolcanism, an informal term representing the eruption of $1000\,km^3$ or more in a single event. Notably, the Kiladze feature bears a structural and geologic resemblance to one of these calderas, Siloe Patera, within the Arabia Terra region (Figure 4). J. R. Michalski & J. E. Bleacher (2013) characterized Siloe Patera as a deeply nested depression resulting from episodes of collapse. The patera has been described as a steep-walled depression associated with arcuate scarps and faults, with structural characteristics comparable to those of Kiladze.

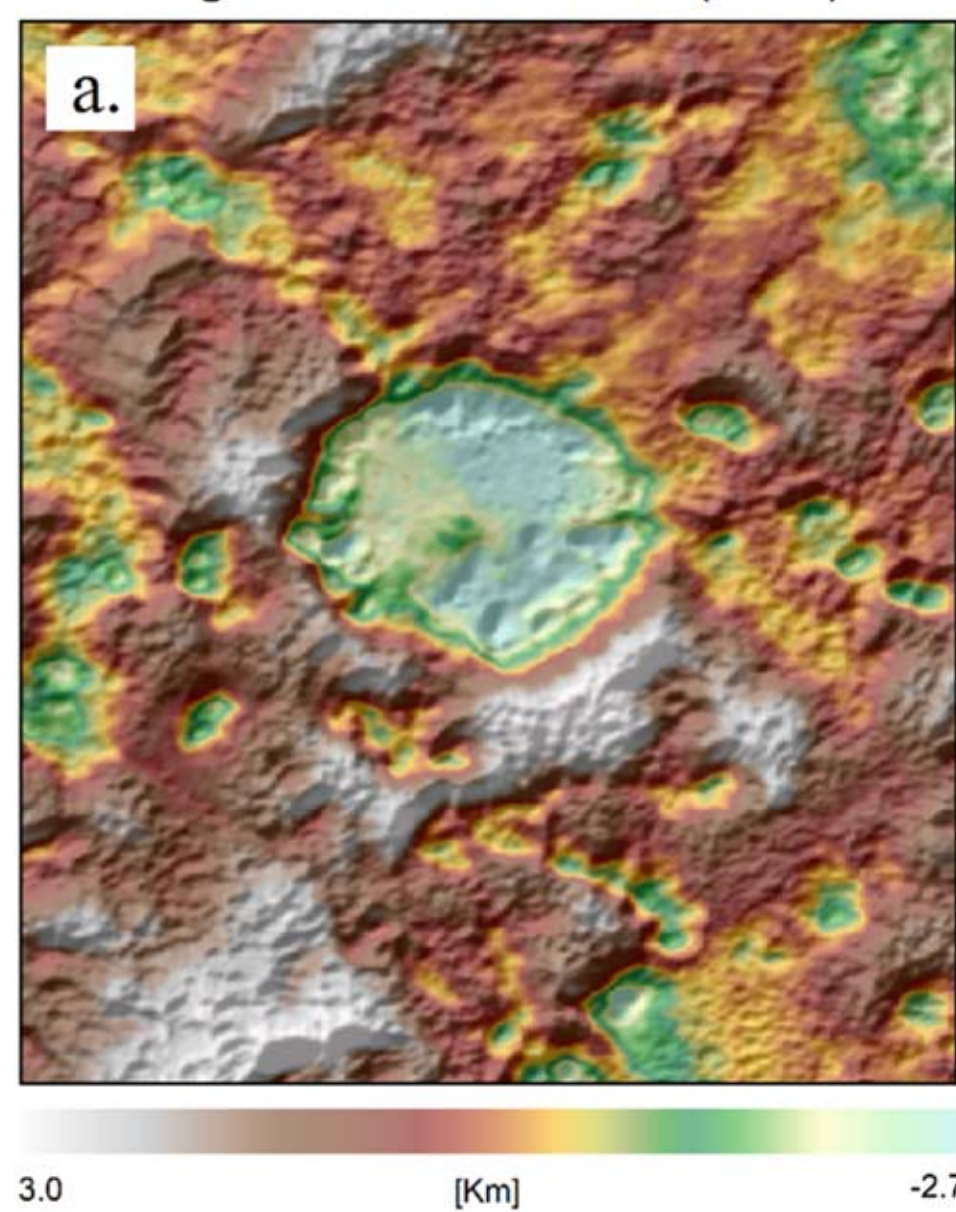

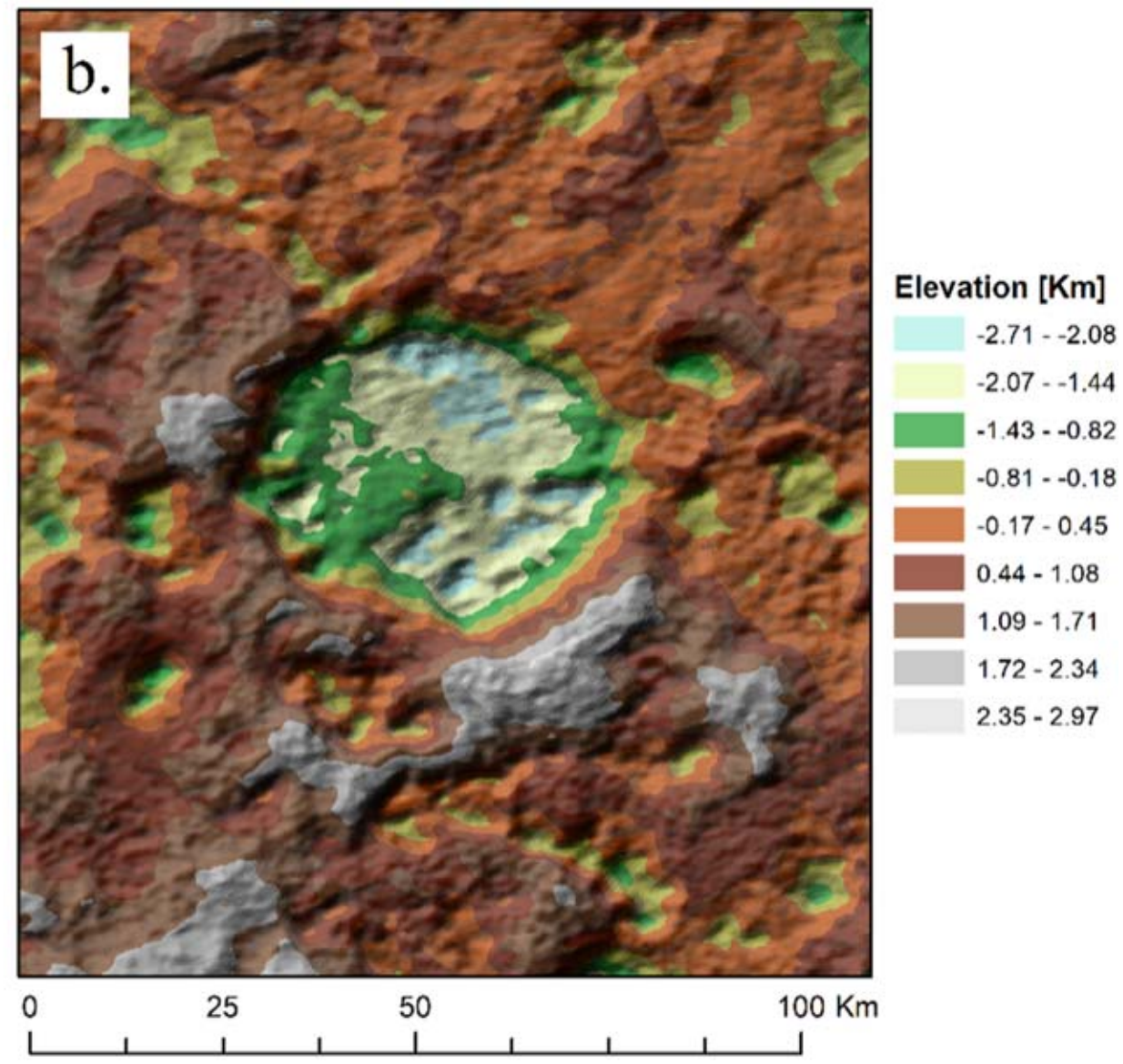

**Figure 3.** (a) Two-dimensional visualization of the DEM and (b) a TIN contour map at the Kiladze area and its surroundings. The DEM (P. M. Schenk et al. 2018) was derived from combined observational data from the New Horizons MVIC and LORRI instruments. The DEM was produced at 100–300 m vertical and 300–800 m spatial resolution, and with a stereo height accuracy from 100 to 1500 m. The data product is peer reviewed and certified for further scientific use for mission data analysis. Note that a 3D visualization of the base map and the TIN can be found in Figure 6.

The base map of Kiladze was produced from the mosaic of highest-resolution images from the LORRI and MVIC instruments (P. M. Schenk et al. 2018). We have made depth transects across four chords in the DEM, all intersecting on the rise on the caldera floor (Figure 5). The highest point near the center stands ∼1000 m above the lowest part of the floor.

As noted in Section 2, this region of Pluto displays substantial tectonization, as manifested in normal faults, troughs, and graben, with surface collapse instigated by such tectonism having produced many of the aligned pit chains. An impact into a preexisting, heavily fractured terrain might potentially expose subsurface conduits through which subsurface volatiles responsible for the distinct spectral signatures in this region could escape via cryovolcanism. As such, we first consider the possibility that Kiladze may be an impact crater. Impact craters can be broadly categorized into two types—simple and complex craters. A simple impact crater typically has a bowl-shaped morphology, while complex crater morphologies commonly have relatively broad, flat, shallow floors, terraced walls, central topographic peak(s), and scalloped rims composed of nested arcuate scarps (e.g., R. J. Pike 1980; S. J. Robbins et al. 2021). We do not see a typical central peak in Kiladze (S. J. Robbins et al. 2021). For Pluto, the simple-to-complex crater transition, based on crater morphology, is approximately 11–12½ km. From a morphological study of many complex and simple craters, S. J. Robbins et al. (2021) proposed an equation for the depth/diameters relationship for complex craters on Pluto as

$$d = \alpha \cdot D^{\beta}, \tag{1}$$

where $d$ is the depth, versus $D$ is the diameter, and the constants $\alpha = 0.346 \pm 0.025$ and $\beta = 0.546 \pm 0.025$. Using this equation, for the ∼44 km average diameter of a crater, the depth for Kiladze should be 2.74 ± 0.50 km. Kiladze can, in principle, be classified as a degraded complex-type impact crater. However, if it indeed represents a degraded impact crater with patterns of secondary craters, where erosion or subsequent melting has removed the typical complex impact features, including an elevated rim and a conspicuous central peak, then we would anticipate a significantly shallower depth. Erosion by cycles of evaporation and deposition (e.g., T. Bertrand et al. 2019), as well as the wind that appears to transport material across some regions of Pluto, would likely have removed the raised rim and central peak. Eroded materials accumulating on the crater floor would result in a reduced depth. Furthermore, geological resurfacing, mass wasting, viscous relaxation of topography, etc. (K. N. Singer et al. 2019), if occurring in that area after the crater formation, could also contribute to a reduced crater depth.

Slumping of the southwest wall into Kiladze (due to caldera collapse) appears to have formed much or all of the smooth-textured floor, with the central rise extending several hundred meters above the mean level (Figure 5). Kiladze has an average depth, measured from the crater rim to the crater floor calculated using the elevation models (P. M. Schenk et al. 2018), of >3 km, while at its lowest points it is >4 km (Figure 5), and it has an area of 1587 $km^2$ (R. A. Beyer & M. Showalter 2021). If it were an impact crater, then certainly it must be a degraded complex crater, considering the morphology of the features. For a degraded impact crater the size of Kiladze, the depth should be substantially shallower than 2.74 ± 0.50 km, for the reasons mentioned above. Even if we consider the upper limit of this number, the maximum depth will be 3.24 km, while the average depth is more than 3 km, and some parts are deeper than 4 km, as seen in the transects (Figure 5).

We compare a similar-sized (diameter ∼46 km) complex crater named Giclas (centered at 39°.72N and 201°.86E) located in the same region, east of the Sputnik Planitia and a few degrees northwest of Kiladze (Figure 6). Giclas crater and its surroundings show no evidence of exposed water ice. The base map and elevation data (Figures 6 and 7) show that Giclas has a prominent central peak, different from the central part of Kiladze, and has a flat, smooth floor. S. J. Robbins et al. (2021) noted the absence of a bowl shape, but the presence of a central peak, flat floor, and scalloped rim. It is thus categorized as a complex crater. However, unlike the Kiladze feature, collapsed structures in the floor of Giclas are not evident. Based on the depth/diameter relationship from S. J. Robbins et al. (2021) in Equation (1), this ∼46 km diameter crater should have a depth of $2.80 \pm 0.50$. The calculated mean rim–floor depth of Giclas is estimated to be 2.41 km (S. J. Robbins et al. 2021; Figure 7), which is expected for a complex crater with a history of degradation after its formation. The slightly shallower depth may result from erosion and subsequent melt deposition as well as other causes. In contrast, rather than having a shallower elevation than estimated depth, Kiladze shows the opposite, having a higher rim–depth elevation. As noted here, the mean rim–floor elevation is >3 km, and based on average values from the transects (see Figure 5), the highest point of the rim to the lowest point of the floor is ∼4.2 km, which is somewhat deeper than the estimated depth by Equation (1). This substantiates the assertion that Kiladze is less likely to be a degraded impact crater, and its geologic structure may have formed through alternative mechanisms or a combination of such processes. It is a possibility that the original impact structure was subsequently modified by pits, as the region shows abundant evidence of modification through the formation of clusters and chains of pits, as illustrated in Figure 2. However, the dissimilarities in surface composition between the Kiladze feature and the nearby pitted areas (Figure 1(d)) suggest that Kiladze may have a different origin. If the feature were an impact structure that was later modified by pits, the nearby pitted areas would be expected to exhibit a similar composition. While the yellow cluster (class 6, shown in Figure 1(d)) corresponds to the Kiladze structure, and its average spectra reveal a signature of ammoniated water ice, the nearby pitted areas display $CH_4$ ice that is depleted in ammoniated features (A. Emran et al. 2023a).

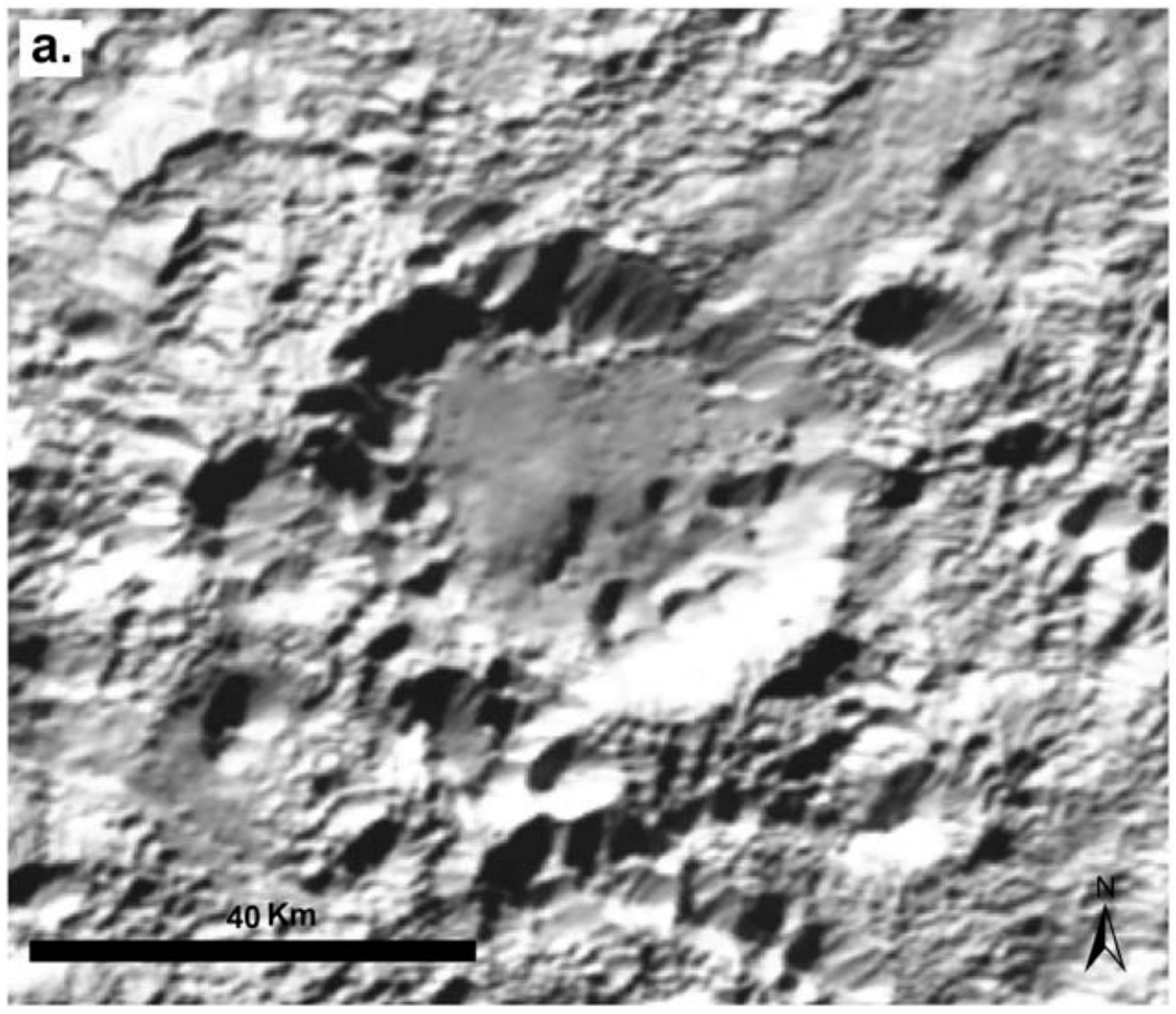

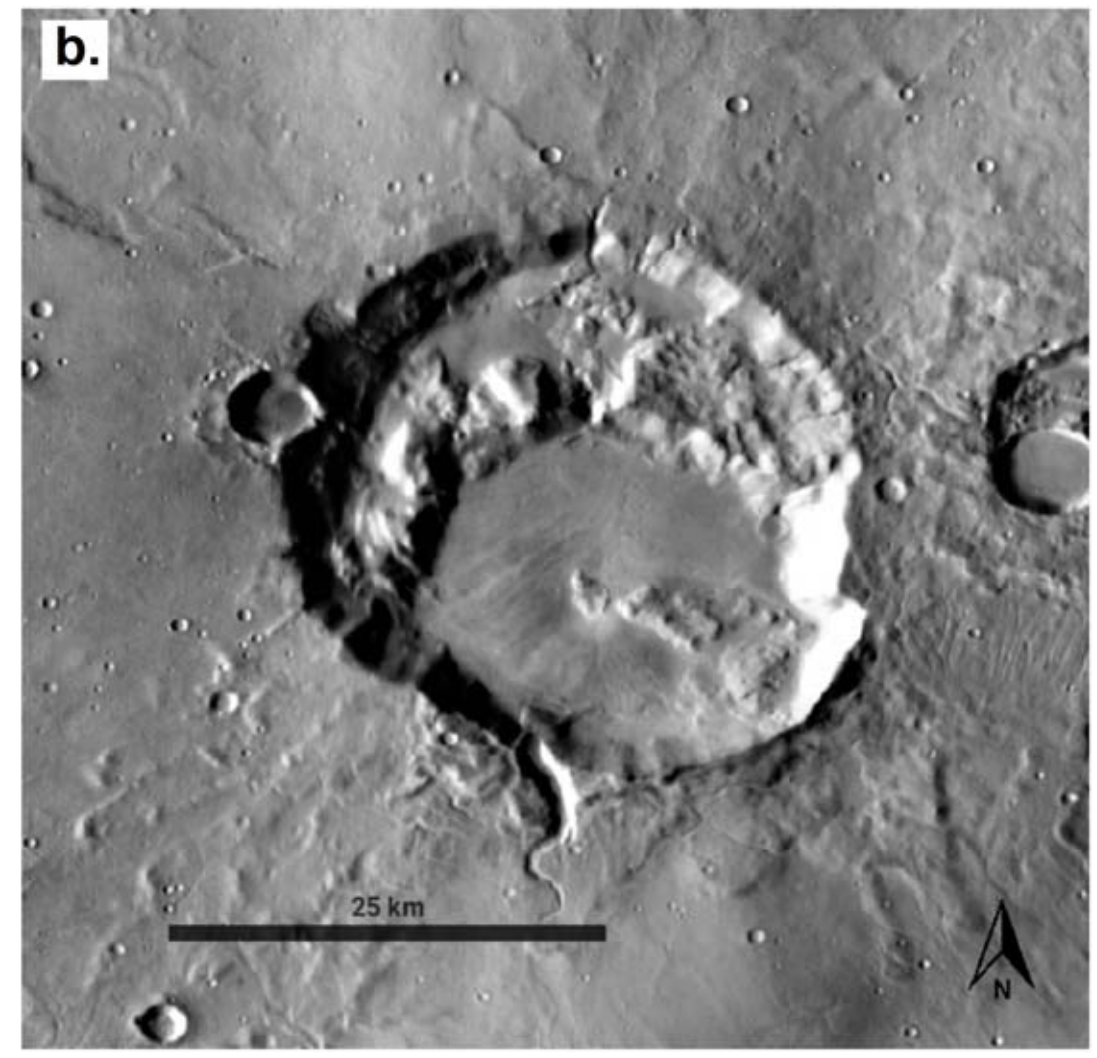

**Figure 4.** Geologic resemblance of calderas on Pluto and Mars. (a) The panchromatic base map (P. M. Schenk et al. 2018) of the Kiladze structure on Pluto, as observed by the New Horizons flyby in 2015 July, and (b) a panchromatic image of Siloe Patera within the Arabia Terra region on Mars, as observed by an instrument on board the Mars Reconnaissance Orbiter.

With all factors taken together, and in view of the asymmetry and slumped morphology (supported further by the compositional evidence), we suggest that Kiladze is less likely to be a degraded impact crater, thus implying that it may have formed through alternative mechanisms or a combination of such processes. On the basis of all the criteria noted above, Kiladze is clearly a complex structure, unlike a typical impact crater in its current form, thus lending weight to our interpretation of it as a caldera complex.

## 5. Volume of Kiladze

The amount of collapse for the Kiladze feature is estimated from the volume calculation of the depression using DEM data (P. M. Schenk et al. 2018). In this instance, the volume of the depression of Kiladze was calculated from the reference 0 km elevation level. Although there are higher elevations than the reference level, we choose 0 km since the 0 km contour connects together around the depression and defines a reference level for a calculation of the crater volume (Figure 8). Thus, the volume calculation measured here from the 0 km reference line is the minimum volume of the depression and is estimated to be $1673 \text{ km}^3$. Note that the volume estimate provided here represents the total void space beneath the 0 km reference line, assuming that the topography has remained unchanged since the occurrence of the collapse event. However, we acknowledge that this assumption may not hold true, given the dynamic nature of surface features on Pluto. Factors such as mass wasting, sublimation, and deposition of volatiles can potentially alter the original topography of the feature since its formation. Still, this estimate offers an approximate indication of the amount of material that has been displaced or removed from the structure as a result of one or more eruptions, leading to caldera formation.

In order to evaluate the volume of the collapse, calorimetric estimates were calculated for a hypothetical removal of subsurface material (i.e., cryomagma material) that would have been required to remove a volume of ice equal to the volume at Kiladze (H. Björnsson 1983; C. J. Ahrens & V. F. Chevrier 2021). For a removal of subsurface cryomaterial, we estimate that the material is a water ice-ammonia mixture, as proposed or observed at other Pluto tectonic features (D. P. Cruikshank et al. 2019b, 2019a; C. M. Dalle Ore et al. 2019). We assume that the volume between the depression floor and the surrounding terrain to make the depression was lost during the depression emplacement, and that no prior cavities existed or filled during the collapse. The volume of cryomagmatic material required to collapse this measured volume at Kiladze, assuming pure water ice at the surface, $V_c$, is (e.g., M. T. Gudmundsson et al. 2004)

$$V_c = (p_i L_i V_i)(p_c C_c \Delta T)^{-1}, \qquad (2)$$

where the subscript $i$ indicates the surface water ice, $c$ indicates cryomagma, $\rho$ is the density (mean ice mantle density 950 kg m$^{-3}$; F. Nimmo et al. 2017; C. J. Bierson et al. 2018), and $L$ is the latent heat of water ice (333 kJ kg$^{-1}$; S. Kamata et al. 2019). From J. S. Kargel (1998), the density of an ammoniated water-ice cryomagma is 980 kg m$^{-3}$ (for an estimation of $\Delta T = 136$ K). The specific heat ($C_c$) for a pure ammonia-water mixture at ∼40 K is 8.494 kJ kg$^{-1}$ K$^{-1}$ (R. Overstreet & W. F. Giauque 1937). From Equation (2) and the estimated volume of Kiladze at 1673 km$^3$, the cryomagmatic volume needed for formation by depression would be 467.5 km$^3$. We note that we neglect heat loss to warming of any icy debris due to unknown temperature gain and ice-mixing ratios, and that the asymmetric structure of Kiladze would also imply turbulence in the removal stage of subsurface cryomagma, possibly leading to subsequent pitting in the vicinity. As points of comparison of overall size, several large calderas are recognized on Earth. Among the largest are La Pacana (Chile) 60 × 35 km, Pastos Grandes (Bolivia) 50 × 40 km, Awasa (Ethiopia) 40 × 30 km, Yellowstone (USA) 85 × 45 km, and Long Valley (USA) 32 × 17 km.

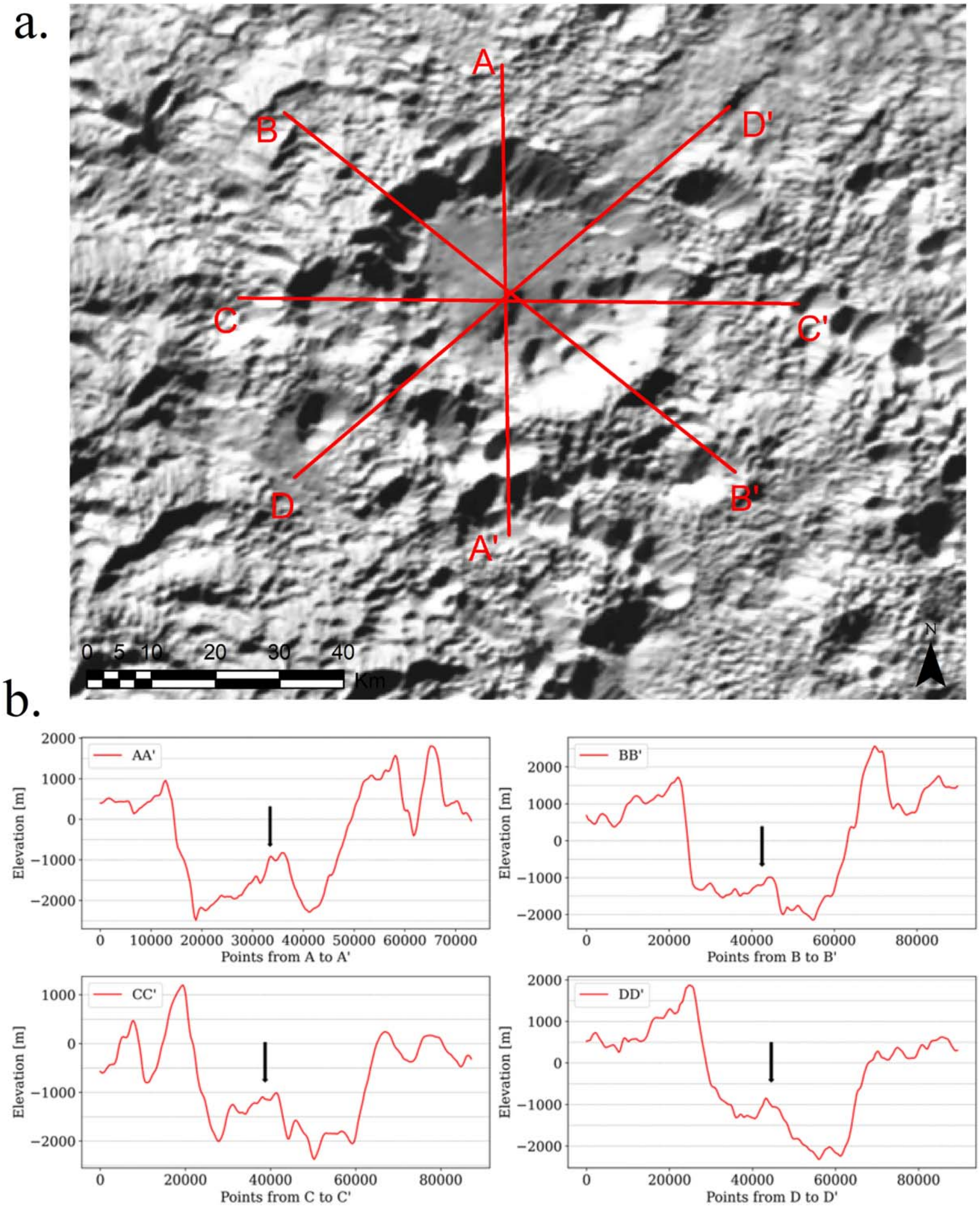

**Figure 5.** (a) Transects to extract the elevation profiles of the Kiladze area are shown in the upper panel. The transects intersect at the peak of the central rise in the smooth terrain. (b) Elevation profiles for the four transects are in the lower panel, with arrows indicating the rise in the smooth-textured material presumed to have slumped into the crater from a collapse of the southwestern wall. The vertical scale is in meters. AA′, BB′, CC′, and DD′ are the four transects.

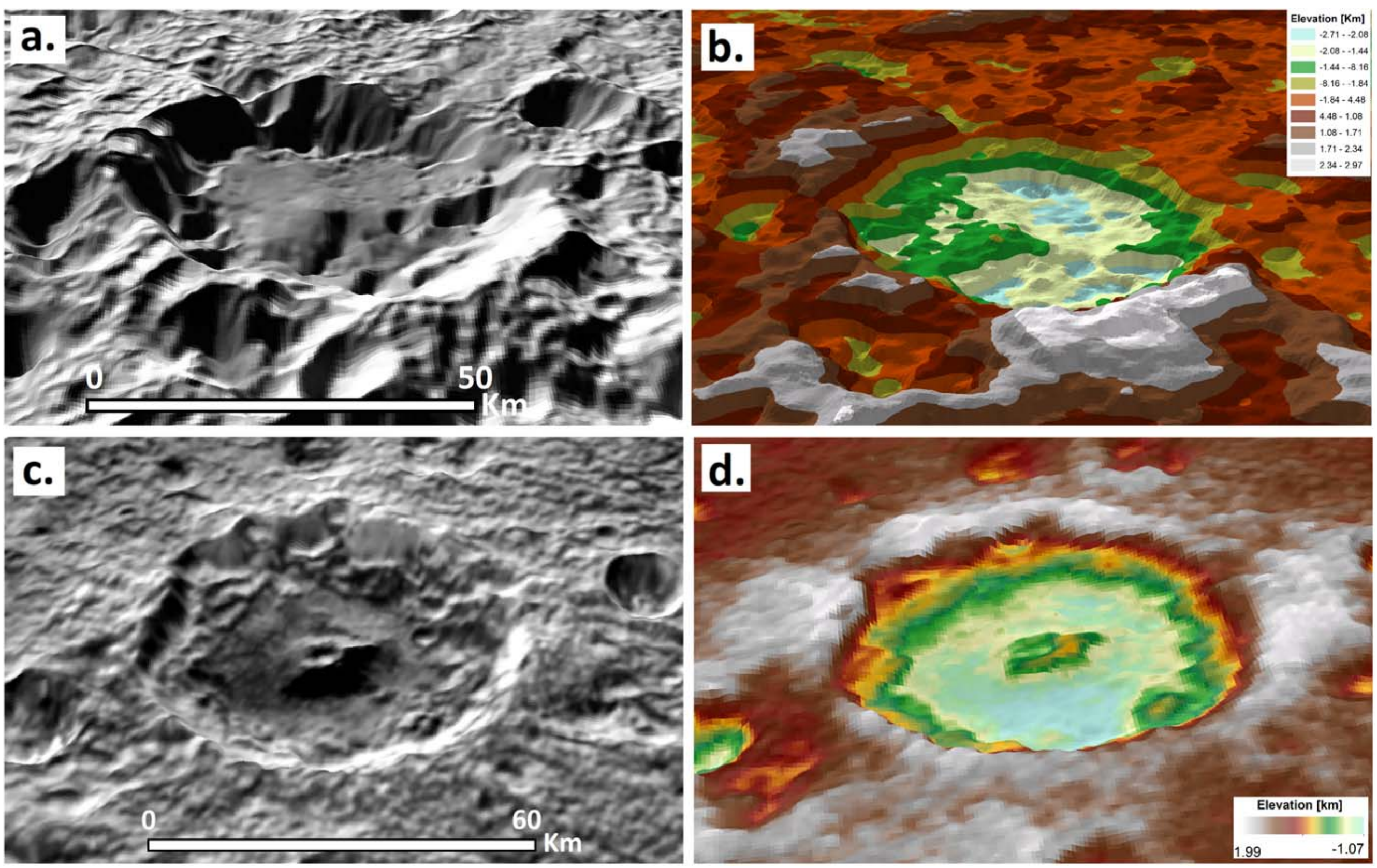

**Figure 6.** Comparison of Kiladze and the similar-sized Giclas crater on Pluto. A 3D visualization of the base map (a) and the TIN map (b) at the Kiladze area and the base map (c) and a DEM (d) at Giclas crater and its surroundings. The 3D visualization of the monochromatic base map at 300 m pixel$^{-1}$ spatial resolution (P. M. Schenk et al. 2018) was rendered vertically to elevation model data and the TIN.

## 6. A Scenario for the Formation of the Kiladze Complex

We have made the case above for the relative youth of Kiladze and the $H_2O$-rich terrain in comparison with adjacent and nearby regions on the planet. In view of that conclusion, a mechanism for the formation of the crater and its surroundings is needed. The most straightforward explanation is that the $H_2O$, carrying a weak component of an ammoniated salt or hydrate, was emplaced by one or more eruptions of a cryomagma through faults, and deposited on the surface. On other icy bodies, there could be substantial reservoirs of subsurface water, including ammonia (R. M. Lopes et al. 2013), as has been observed in other examples of icy depressions on bodies such as Triton and Ganymede. These bodies exhibit scalloped depressions, most probably arising through some removal of volatiles (and sublimation; W. B. McKinnon et al. 2001; N. A. Spaun et al. 2001). Proximal fissures and graben show a commonality among collapse features associated with caldera formation (or any feature with the removal of some subsurface material; V. Acocella 2006; L. Michon et al. 2009; A. Gudmundsson 2016). We note that Kiladze's elongated planform, with the long axis being aligned parallel to the northwest–southeast trend of faults and collapse pit chains in the region, bears a strong resemblance to the large, ovoid depressions in eastern Belton Regio (previously known as Cthulhu Macula) that coincide with the hemispheric-scale ridge-trough system and its constituent faults (P. M. Schenk et al. 2018), the north-northeast–south-southwest trend of which their long axes are parallel.

C. J. Ahrens & V. F. Chevrier (2021) investigated Hekla Cavus, the northernmost of these depressions in Belton Regio, and also determined that its morphology is different from those of impact craters on Pluto. They estimated that it has a volume of 5617 km$^3$ of removed ice, which they interpreted to suggest a collapse origin stemming from a cryovolcanic disturbance related to the major tectonic uplifting of the icy lithosphere associated with the formation of the ridge-trough system. We also note that most volcanic structures on rocky planetary bodies are primarily formed at the interaction of differing compositional settings, like mantle (subsurface) magma and crustal solids (W. Bosworth et al. 2003). Although Kiladze is in a structurally different environment than the Virgil Fossae and Viking Terra exposures of $H_2O$, the basic eruptive mechanism could be similar, although recurrent. In view of the large scale of the Kiladze crater, diverse morphology of the floor and deep collapse pits, and the extent of the surrounding $H_2O$ ice emplacement, this structure could plausibly be interpreted as a cryovolcanic caldera complex.

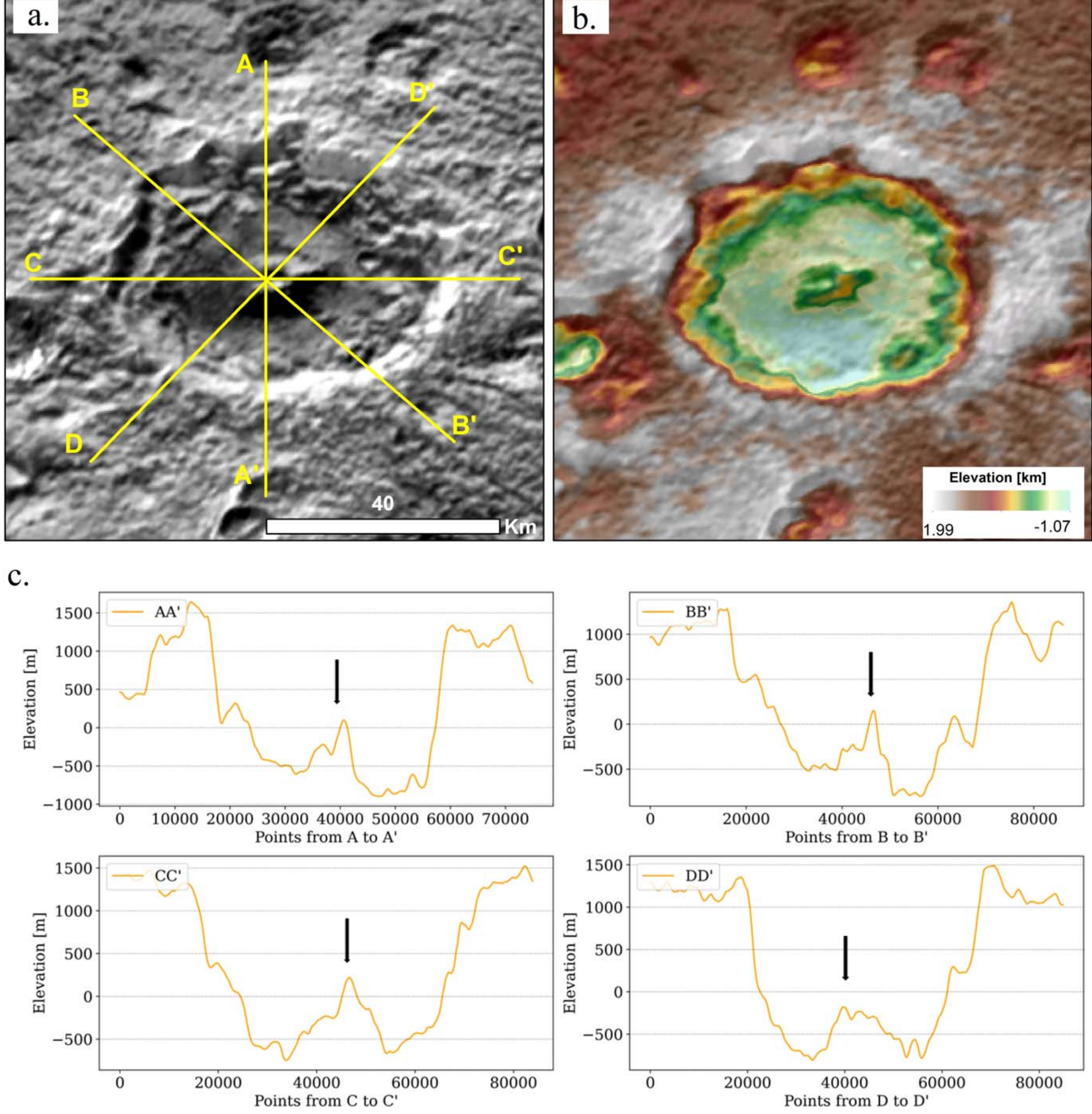

**Figure 7.** Giclas crater with a diameter of ∼46 km. (a) The highest-resolution base map image of the region around the crater. Yellow transects to extract the elevation profiles of the crater, shown in (c). The transects intersect at the peak of the central rise. (b) Two-dimensional visualization of the DEM of Giclas crater. (c) Elevation profiles for the four transects on (a), with arrows indicating the central peak of the crater. Vertical scale in meters. AA′, BB′, CC′, and DD′ are the four transects.

The formation of the caldera wall is probably the signature of a ring fault system in which one or more recurring eruptions occurred, each punctuated by collapse and the formation of the deep pits now found in and around the basic circular wall structure. These pits may represent a combination of collapse and explosive events similar to volatile-driven explosive pits found on Earth and Vesta or Ceres (H. G. Sizemore et al. 2017; S. S. Hughes et al. 2018; T. Michalik et al. 2021). At the same time, water-rich effusions along faults near the ring could spread over the entire region surrounding the caldera in one or more episodes, resulting in the distribution now seen. It is also possible that very violent explosive eruptions of the cryomagma covered the region, perhaps not unlike the huge eruptions at the Yellowstone and other terrestrial resurgent calderas, as deduced from the geological records at these sites. The asymmetry in the smooth-textured component on the caldera floor may represent asymmetric eruptions and suggests a downward motion of slumped material to the depression floor as a postformation process. This is similar to a mass-wasting lobe on the floor of Hekla Cavus, observed by C. J. Ahrens & V. F. Chevrier (2021), which they regarded as an indication of displacement and collapse of material (rather than simply sublimation or impact).

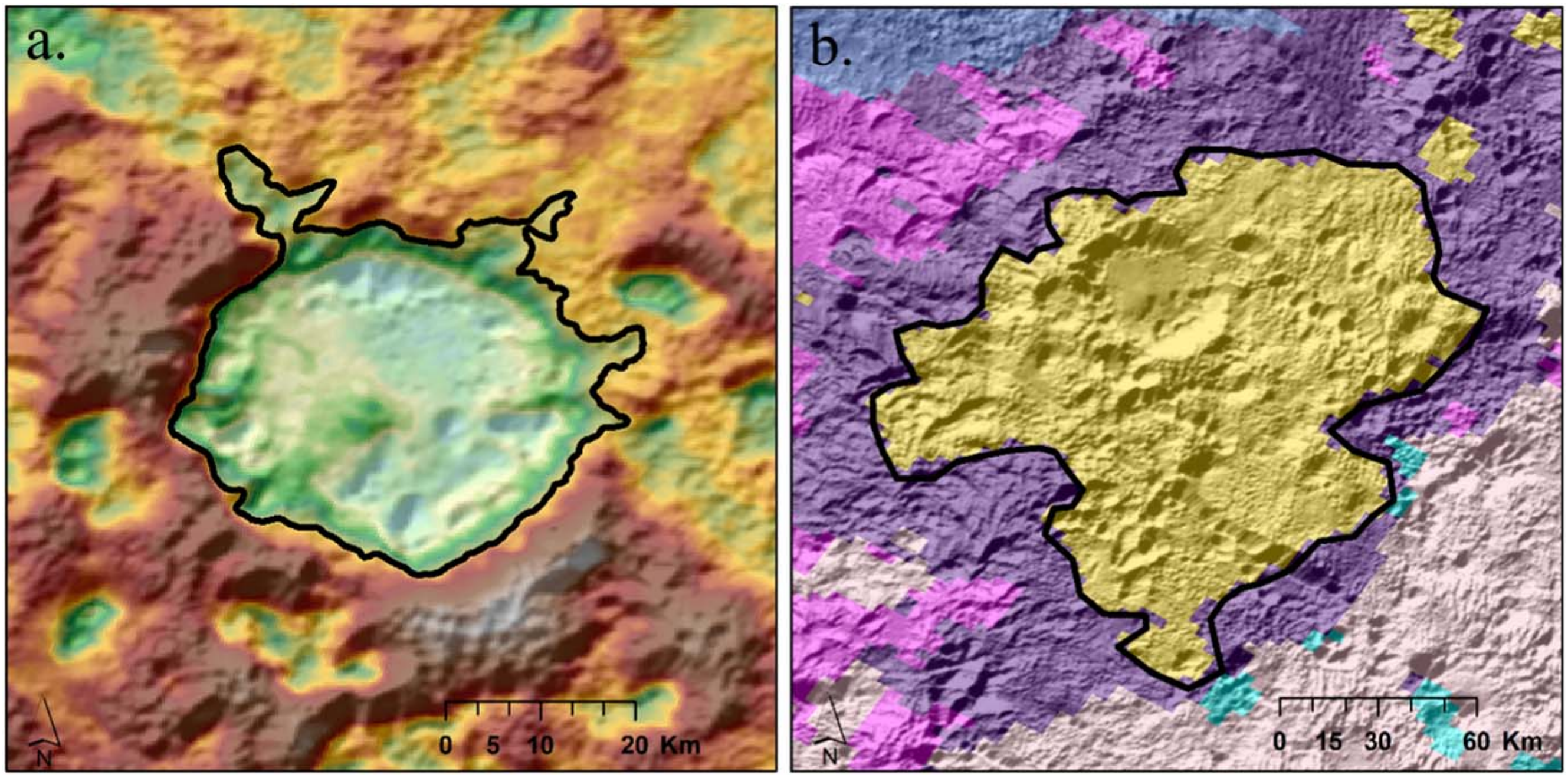

**Figure 8.** (a) The 0 km contour (black line) overlaid on the elevation model used for calculation of the caldera volume. (b) The proposed outline (with much stronger ammoniated water ice) of deposit (black line) overlaid on the compositional map modified from A. Emran et al. (2023a). The differences in the orientation of the outline and compositional units in this figure vs. Figure 1(d) are due to the image projections used (simple cylindrical vs orthographic, respectively).

Such asymmetry suggests that this structure is in the category of "trapdoor" calderas; similar examples are found on the Moon and on Mars (V. Acocella 2006). In view of the morphology, greater depth than other similar craters on Pluto, and resemblance to "plains-style caldera complexes" of Mars, we assert that Kiladze is very likely to be a cryovolcanic caldera feature. The minimum estimated cryomagmatic volume of 467 $km^3$ at Kiladze was calculated by considering the minimum depth of the depression; however, considering an average depth of 3 km, the volume would be more than 2 times higher. The exposures of water-ice cryolava on the surface likely extended well beyond the currently estimated margins. Accounting for one or more eruption events that displaced material suggests that the actual volume is significantly higher than the minimum estimate, potentially approaching 1000 $km^3$, which, by some definitions, characterizes a supervolcano.

## 7. Conclusions

The Kiladze structure in Pluto's Hayabusa Terra is a ∼44 km diameter depression whose outline is distorted in accord with the local structural trends, and whose walls are largely defined by clusters of pits and slumps. Collapse pits punctuate much of the local terrain, in addition to numerous northwest–southeast trending normal faults. Kiladze itself and the landscape—as much as ∼100 km in several directions—display the near-infrared signature of water ice, as seen in New Horizons spectral maps, standing in strong contrast to the superficial methane ice layer that dominates the surrounding surface. A substantial volume of water-ice-rich material likely occurs beyond the region where it dominates, but is dispersed in patches too small to be resolved with the spectral maps obtained by New Horizons. Water ice is regarded as the bedrock of Pluto's crust, and its persistent appearance at Kiladze, despite the accumulation of atmospheric aerosols formed in the planet's atmosphere that have fallen to the surface over the history of the planet, supports the view that it is a relatively young exposure. In consideration of the calculated rate of aerosol fallout and accumulation, the exposed water ice at Kiladze is of the order a few million years in age or possibly somewhat older, but much less than the age of Pluto's overall surface. In consideration of the size, structure, composition, and youth of Kiladze and its surroundings, we suggest that this region is a cryovolcano with a caldera structure, having a history of one or more eruptions ejecting $10^3$ $km^3$ of cryolava, and possibly an unknown number of eruptions of a smaller scale.

## Acknowledgments

Part of this research was carried out at the Jet Propulsion Laboratory, California Institute of Technology, under a contract with the National Aeronautics and Space Administration (80NM0018D0004). We also acknowledge support from the New Horizons mission, NASA Goddard Space Flight Center and the University of Maryland, and NASA's Planetary Data Archiving, Restoration, and Tools Program (80NSSC19K0423). We thank F. Nimmo, S. A. Stern, and A. D. Howard for helpful comments on earlier drafts of this paper.

## Data Availability

The derived data (I/F) products of LEISA scenes generated by the New Horizons Pluto encounter surface compositional science theme team, higher resolution base map data product, and elevation models derived from combined data MVIC and LORRI images are publicly available at NASA's Planetary Data System: Small Bodies Node at https://pds-smallbodies.astro.umd.edu/ (A. Stern 2018).

## Appendix
## Clustering Analysis of LEISA Data

For the clustering analysis to identify different compositional units in the Kiladze area, we used the principal component-reduced Gaussian mixture model (PC-GMM; A. Emran et al.

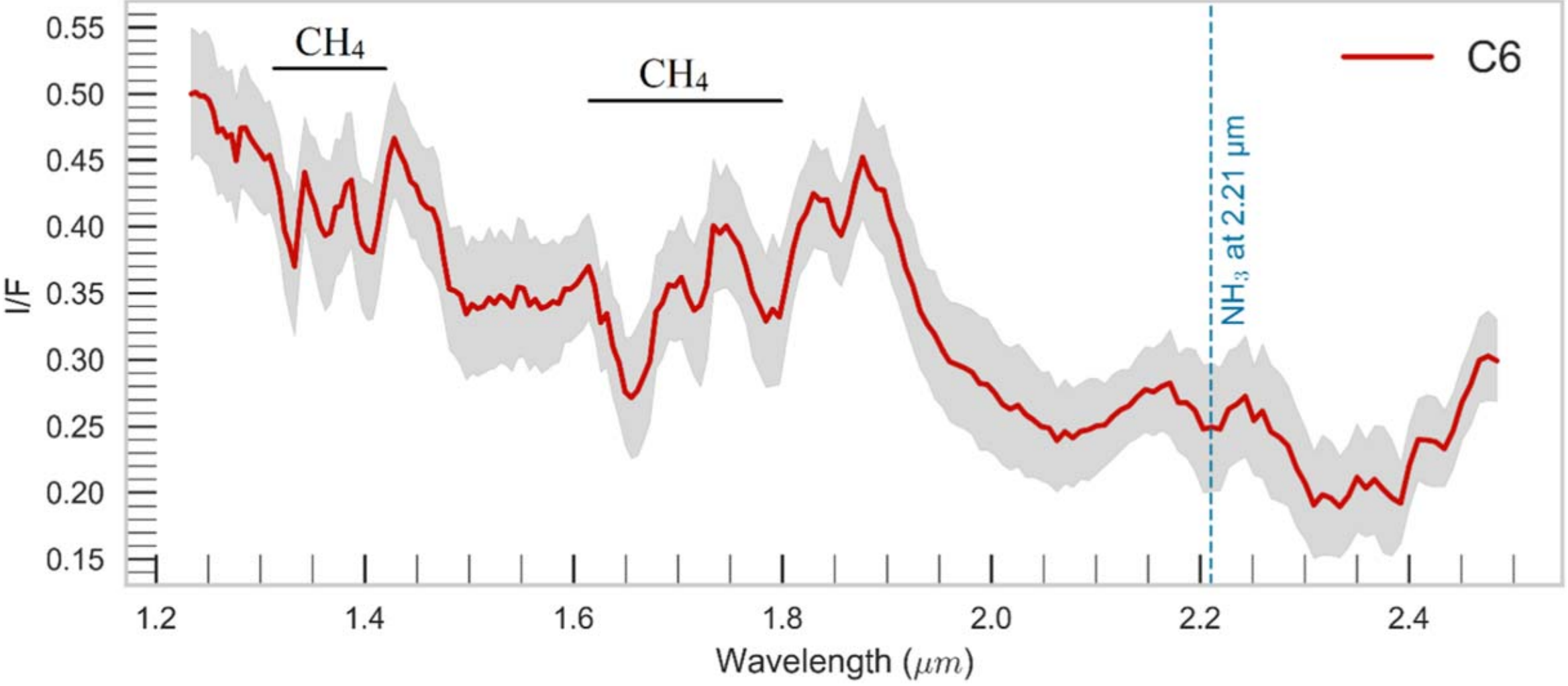

**Figure A1.** The mean (red line) $\pm 1\sigma$ standard deviation (gray shade) of radiance factor (I/F) spectra of Kiladze and its surroundings. Note that cluster number six (C6) represents the yellow-coded region in Figure 1(c)—corresponding to Kiladze and its surroundings. The spectrum shows the signatures of $H_2O$ ice absorptions bands at ∼1.5 and ∼2.0 $\mu$m, $CH_4$ bands at 1.3–1.4 $\mu$m and 1.65–1.8 $\mu$m, and a broad absorption at ∼2.21 $\mu$m—suggesting a presence of potential ammoniated materials. The figure was adopted and recreated from A. Emran et al. (2023a) with permission.

2023b), with data from the Linear Etalon Imaging Spectral Array (LEISA). LEISA is a mapping spectrometer on board the New Horizons spacecraft, and covers the wavelength range 1.25–2.5 $\mu$m, divided into 256 spectral channels (D. C. Reuter et al. 2008). With the PC-GMM routine, we first reduced the LEISA data dimension using principal component analysis, followed by an implementation of the multivariate GMM to the reduced data (A. Emran et al. 2023a, 2023b). Accordingly, the PC-GMM resulted in different clusters of surface compositional units in the Kiladze area, as shown in Figure 1(d). The corresponding average spectra for each cluster can be found in A. Emran et al. (2023a). The yellow cluster (class 6) shown in Figure 1(d) corresponds to the Kiladze structure, and the average spectra for this class showed a signature of the ammoniated water-ice feature (Figure A1; A. Emran et al. 2023a).

## ORCID iDs

A. Emran https://orcid.org/0000-0002-4420-0595
D. P. Cruikshank https://orcid.org/0000-0002-0541-5569
C. J. Ahrens https://orcid.org/0000-0003-1574-112X
O. L. White https://orcid.org/0000-0003-0126-4625